# Low-Energy Switching of Antiferromagnetic CuMnAs/ GaP Using sub-10 Nanosecond Current Pulses

K. A. Omari, L. X. Barton, O. Amin, R. P. Campion, A. W. Rushforth, P. Wadley and K. W.

Edmonds

*School of Physics and Astronomy, University of Nottingham, Nottingham NG7 2RD, United Kingdom*

**Abstract**

The recently discovered electrical-induced switching of antiferromagnetic (AF) materials that have spatial inversion asymmetry has enriched the field of spintronics immensely and opened the door for the concept of antiferromagnetic MRAM. CuMnAs is one promising AF material that exhibits such electrical switching ability, and has been studied to switch using electrical pulses of length millisecond down to picosecond, but with little focus on nanosecond regime. We demonstrate here switching of CuMnAs/GaP using nanosecond pulses. Our results showed that in the nanosecond regime low-energy switching, high readout signal with highly reproducible behaviour down to a single pulse can be achieved. Moreover, a comparison of the two switching methods of orthogonal switching and polarity switching was done on same device showing two different behaviours that can be exploited selectively for different future memory/processing applications.

1. Introduction

Writing of magnetic information using spin-polarized currents, via current-induced spin transfer and spin-orbit torques, has been key to the development of magnetic random-access memory (MRAM) technologies[1, 2]. Recently, current-induced switching has been predicted and demonstrated in antiferromagnetic (AF) materials, enriching the field of AF spintronics[3-19]. With the ability to manipulate AF domains electrically, memory-based spintronic applications can be developed that utilize the terahertz dynamics[8] and multilevel response[7] of the AF order, while also producing no stray magnetic fields and being insensitive to external fields.

Current-induced switching of AF materials relies on generation of a spin-orbit torque with the same handedness on each magnetic sublattice. This can in principle be achieved in AF crystals with a sublattice inversion asymmetry, including CuMnAs and $Mn_2Au$[4-11, 15, 20], as well as in AF/heavy-metal multilayer structures[12-14]. In CuMnAs, switching of AF domains between metastable biaxial easy axes has been demonstrated, with correlation shown between the directly imaged magnetic domain configuration and the electrical readout[6]. The rotation of the AF moments between two perpendicular orientations can be realized by applying current pulses along two orthogonal axes in a multi-contact device, and can be electrically detected using anisotropic magnetoresistance (AMR), *i.e.* different electrical resistivity for current flow parallel and perpendicular to the AF spin axis[5]. Patterned CuMnAs devices have been shown as a proof-of-concept for a simple memory cell that can be integrated with existing CMOS technology[7].

More recently, another form of switching in CuMnAs was demonstrated, by reversing the polarity of a current pulse applied between a single pair of electrical contacts[9]. The resulting spin-orbit torque can cause motion of domain walls separating the two orthogonal domain populations, resulting in measurable changes of the AMR. This polarity-dependent switching method offers simpler device structures and potentially also lower switching currents compared to the orthogonal switching method.

While electrical switching of CuMnAs/GaP has been previously investigated using ms[5,9] and ps[7] pulses in details, the ns pulsing regime remains not yet fully studied. This regime is relevant and significant for a low-energy switching regime that can be easily integrated with standard microelectronics, as opposed to the relatively slow ms regime, or the ps regime that would require either optical switching techniques or highly sophisticated electronics.

In this paper, we investigate the switching behaviour of a CuMnAs/GaP device in the nanosecond (ns) regime. Both switching methods of orthogonal switching that shows a slow decay of signal, and polarity switching method which shows a highly non-volatile switching behaviour, are explored on same device using our setup.

2. Device Fabrication and Experimental Setup

A 46nm layer of CuMnAs was grown using molecular beam epitaxy (MBE) on a 2" substrate of single crystal GaP (100)[21,22]. Prior to growing the CuMnAs layer, the substrate was annealed at high temperature under P flux to remove the surface oxide layer. A 100nm homo-epitaxial GaP buffer layer was grown to ensure a smooth surface for the CuMnAs. The hetero-epitaxial CuMnAs layer was then grown, followed by a 2.5nm capping layer of Al to prevent surface oxidation. Following the growth process, $2\theta-\omega$ x-ray diffraction scans were performed to confirm that the tetragonal structure was achieved with *c*-axis out of plane, and with the square ab plane of the tetragonal unit cell orientating at a planar 45° offset to the <100> crystal axes of the cubic GaP unit cell. At 46nm thickness, the CuMnAs layer is expected to have a biaxial magnetic anisotropy with in-plane easy axes separated by 90°, as opposed to the uniaxial anisotropy which is typically observed in films thinner than ~20nm[23, 24].

The CuMnAs film was then patterned into eight-arm devices with a central junction of 3.3x3.3µm$^2$ and a 2µm arm width using photolithography and wet chemical etching. Optical micrographs of the device are shown in Fig. 1. Multiple samples were fabricated on a single chip. The chip was mounted on a sample holder (Fig. 1(a)), designed to allow nanosecond current pulses to be driven from the main pulse input terminal via and pair of arms of the device to the grounding output terminals (labelled on the bottom side of the holder with the green grounding line). The current path is selected either by manual toggle switches or by electric relays controlled by an external Arduino controller module.

By activating a toggle switch / relay at one of the four input paths and a toggle switch / relay at one of the grounding output terminals, the direction of propagation of the ns current pulse can be set. Two pulsing configurations were investigated in this work. For orthogonal switching, the pulse is directed alternately along the [$\bar{1}\bar{1}0$] and [$1\bar{1}0$] axes of the CuMnAs film (dashed white arrows in Fig.1(b)). For polarity pulsing, the pulse is directed alternately along [1 0 0] and [$\bar{1}$ 0 0] directions (dashed blue arrows in Fig. 1(b)). In the latter case, an external invertor was connected to the pulse generator via a bridge connection to selectively invert the polarity of the signal before entering the waveguide.

The amplitude and length of each pulse was set to be between 12 V-14 V and 4 ns, respectively, using an Avtech pulse generator. To generate a desired number of pulses, the pulse generator was set to be triggered externally by another function generator that can be automated using a computer software and can produce a square signal of a sharp rising edge of 5V. The two-wire resistance of the devices is typically around 475 Ω. This relatively high resistance with respect to the 50 Ω impedance waveguide leads to some signal reflection as can be seen from the pulse waveform in Fig. 1(a), inset. However, this signal distortion is within tolerable limits as the square shape of the pulse is fairly intact.

In addition to the waveguide paths, the 8 terminals of the device were connected to normal DC circuitry via additional input/output terminal pins. These can be seen on the right and left edge of sample holder in Fig.1 (a). The DC circuitry was primarily used for probing the transverse and longitudinal resistances of the device, $R_{xy} = V_{xy} / I$ and $R_{xx} = V_{xx} / I$, following the application of current pulses. The transverse $V_{xy}$ and longitudinal $V_{xx}$ voltages were continuously measured with a probing current $I$ of 500 µA as depicted in Fig. 1(b). High impedance resistors and high-pass filters were added to decouple the high-frequency circuit and the DC circuit. All pulsing and measurements were performed in ambient conditions.

## 3. Results

*3.1 Orthogonal switching*

In studying the orthogonal switching in the ns regime, the dependence of the transverse resistance on the number of applied pulses was investigated. Moreover, the switching robustness of the device was later tested to assess its reproducibility and stochasticity. To achieve the first objective, ten sets of measurements were performed, each consisting of a total of 30 switching attempts (15 pairs of switches), in which pulse trains were sent alternately across two orthogonal axes with an interval of one minute between each switching attempt. Between each set of measurements, the number of pulses was increased by an additional pulse; such that the first set had a single 4 ns pulse for each switching attempt, and the tenth set had a train of 10 pulses. For sets 2 to 10, the multiple pulses triggered at each switching attempt were separated by 1 µs. The amplitude of each pulse was set at 14 V producing a current of 29 mA, corresponding to a current density of 32 MA/cm$^2$ in the arms of the device.

Fig. 2(a) shows the $R_{xy}$ traces obtained for 5 pulses per switching attempt. After each pulse, the $R_{xy}$ shows either an increase or a decrease depending on the direction of the pulse, consistent with an expected AMR effect due to a transient rotation of antiferromagnetic domains[6, 24]. The $R_{xy}$ relaxes back to an equilibrium value on a timescale of ≈ 20 s. Fig. 2(b) shows the dependence of $\Delta R_{xy}$, defined as the difference between the first and last measured points after each pulse train, versus the number of pulses. The plot shows an asymmetry between the two orthogonal current directions and a non-monotonic dependence on the number of pulses, which may indicate that both the switching and relaxation processes are thermally assisted.

To further investigate the robustness and reproducibility of orthogonal switching in the ns regime, the same switching protocol, with 5 pulses per switching attempt, was run for 100 switching attempts. Fig.3 shows the resulting $R_{xy}$ trace indicating a consistent and highly reproducible switching as suggested by the histogram at inset (further discussion later). The maximum $\Delta R_{xy}$ was calculated to be around 0.1 Ω, and the signal relaxes to its initial value in around 20 s.

As indicated in Fig.2(b), any switching signal in this device was below the noise level for one and two pulses. However, consistent switching using a single pulse was seen in a duplicate device fabricated on the same CuMnAs layer in the same fabrication process. Fig. 4 shows a total of 40 switching attempts (20 switching attempts on each orthogonal axis) using a single 4 ns pulse of amplitude 12V (J=27 MA/cm$^2$). In this device 20 switching attempts were performed along one axis with an interval of one minute between each attempt. Then 20 switching attempts were performed on the second axis to show the reverse effect. This process simulates a simple binary writing of a "leaky" memory state with a single 4 ns electric pulse. A total energy of 1.2 nJ was needed for switching in the active area of 2μmx2μmx45nm of the CuMnAs device. This is orders of magnitude less than energy obtained in the ms regime[7] (where a 5V USB, 50ms pulse and 46mA were used to switch CuMnAs/GaP). The 1.2nJ is also in the same order of magnitude of energy needed by laser for ultra-fast optical switching[25].

*3.2 Polarity switching*

For the polarity switching geometry, first we investigated how changes in $R_{xy}$ correlate with the number of pulses, starting from a single pulse to train of 10 pulses, with each pulse separated by 1 μs. Figure 5(a-c) shows the $R_{xy}$ traces for sets of measurements with 2,3 and 4 pulses per switching attempt. Results showing $\Delta R_{xy}$ versus number of pulses for all the sets of measurements are summarized in Fig. 5(d).

As indicated in Fig. 5(a-c), $R_{xy}$ shows a step-like change between the high and low values depending on the polarity of the switching pulse. $\Delta R_{xy}$ initially increases with increasing number of pulses, and saturates above 4 pulses. $\Delta R_{xy}$ is defined here as the height of the step-like behaviour which resembles the difference in $R_{xy}$ between two consecutive switching events with opposite polarity. In order to compare $\Delta R_{xy}$ between polarity and orthogonal switching, a graph in Fig. 5(d) representing the full step height in orthogonal switching was plotted in same graph (black circles). The step height in orthogonal switching is simply defined to be the change in $R_{xy}$ between the first point of switching of one arm and the first point of switching in the next switching event when the orthogonal arm is pulsed. Moreover, unlike in the case of orthogonal switching where the $R_{xy}$ relaxes back between each pulse train, the results of polarity pulsing indicate no or very slow relaxation of signal at room temperature. Such observation is consistent with previously reported results of polarity pulsing in the millisecond regime[9]. To further investigate the retention time, the device was switched to the high $R_{xy}$ value and left for 15 hours before measuring again. $R_{xy}$ decreased only by 5% over 15 hours (Fig. 5(e)). This slight magnetic decay could be contributed by variations in the ambient temperature as suggested by the increase in $R_{xx}$ (Fig 5(f)). Plots in Fig. 5(d) clearly indicate a much higher $\Delta R_{xy}$ signal obtained for polarity switching compared with orthogonal switching. This suggests that when orthogonal switching occurs, only a partial reorientation of domains occurs at the central region of the device. However, when polarity switching occurs it seems that domain wall movement results in reorientation of a wider area of domains leading to a higher readout signal.

We also investigated the robustness, reproducibility and stochasticity of the polarity switching of the device by continuously switching it for 500 times while maintaining an interval time of 1 minute between each switching attempt. Fig. 6(a) shows the $R_{xy}$ trace for the first 60 switching attempts across 60 minutes. The corresponding $\Delta R_{xy}$ histogram in Fig. 6(b) demonstrates the stochastic behaviour of the readout voltage (see also Fig. S1 in supplementary for the whole of the 500 switching events). It can also be seen in Fig. 6(a) that there is considerable background drift. The origin of this drift is unknown but may be related to incomplete switching due to random pinning / depinning of domain walls, or external factors such as changes in ambient temperature, contact resistances or current paths."

The histogram plot in Fig. 6(b) obtained from the 100 switching attempts indicate a clear difference in the statistical switching behaviour between orthogonal (purple) and polarity (green) switching method. The Gaussian curve fitting for the two histograms indicated a curve width, $wc$, at half the maximum of each histogram to be 0.01 Ω for orthogonal and 0.06 Ω for polarity switching. This indicates that a relatively more deterministic behaviour is seen in the orthogonal switching compared with polarity switching. As discussed above, this suggests that in orthogonal switching the domain reorientation follows the same route in most of the switching attempts, indicating that the switching energy barrier is more well-defined than in the case of polarity switching where a DW can face different pinning energy barriers in each switching attempt. Moreover, the different signal and stability between the two switching methods might also arise from the magnetocrystalline anisotropy for pulsing across different crystal orientations[20].

The relatively small signal obtained from orthogonal switching (compared with polarity switching) suggests that small domain regions are switching only. On the other hand, in the case of polarity switching, the wide distribution of output signal indicates that domain motion in CuMnAs encounters pinning energy landscapes where pinning seems to follow a thermally assisted switching distribution.

4. **Conclusion**

In this paper, we demonstrated the ability of CuMnAs device to switch consistently and reliably using only 1-10 current pulses of length 4 ns. Switching using orthogonal current pulses along the $[\bar{1}\,\bar{1}\,0]$ and $[1\,\bar{1}\,0]$ directions shows a slow decay of the readout signal, while polarity switching for current pulses along the $[1\,0\,0]$ axis shows a highly non-volatile switching behaviour. These results are promising for the development of a robust antiferromagnetic MRAM. Further work is required to establish the role of the magnetic anisotropy in determining the stability of the switched state. With the same medium showing both a "leaky" memory state that can be tuned by temperature[10], and a more stable switching behavior that can be controlled by polarity of the pulse[9], both switching mechanisms can be utilized to perform memory and processing calculations simultaneously.

**Figure 1**

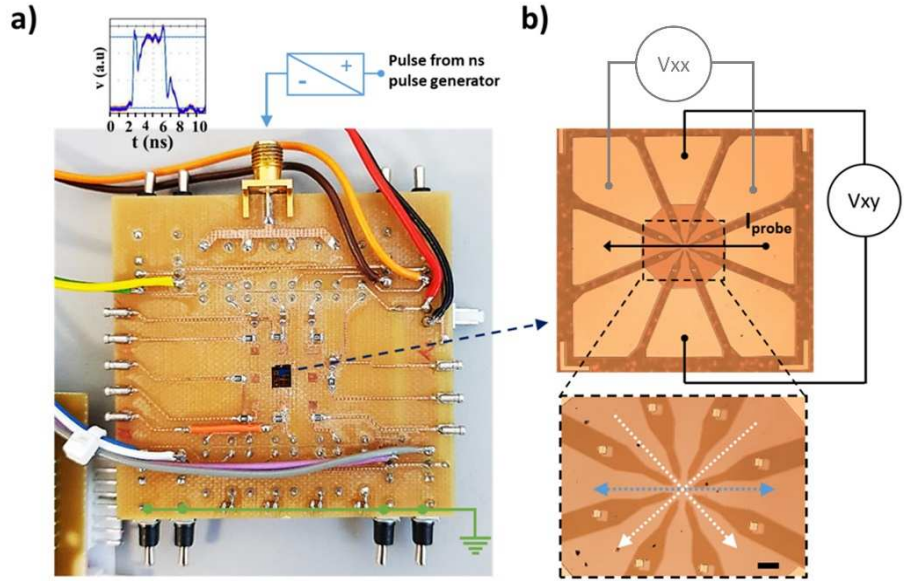

FIG. 1: (a) Photo of the waveguide sample holder showing the controllable toggle switches used to select the current pulse path. The inset above the image shows the shape of the 4ns current pulse. 4 and 3 input pins on left and right side of the holder, respectively, are used to connect probing DC current and voltages. (b): Optical image of the eight-arm CuMnAs/GaP device. The upper image shows the full device with measurement schematic. The lower image shows a close-up of the central junction with dashed arrows indicating the direction of pulsing for orthogonal (white arrows) and polarity (blue arrows). The scale bar is 4μm.

**Figure 2**

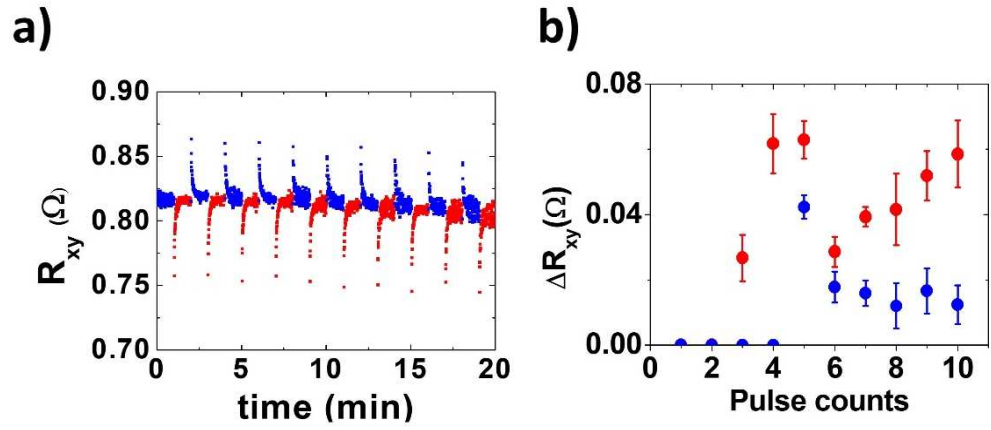

FIG. 2: Readout signal, $R_{xy}$, for orthogonal switching of CuMnAs device. (a): 20 switching attempts using 5 pulses of current 29mA and length of 4 ns. (b): Summary plot of number of pulses versus $\Delta R_{xy}$ for orthogonal switching for pulsing along [$\bar{1}\,\bar{1}\,0$] (red circles) and [$1\,\bar{1}\,0$] (blue circle).

**Figure 3**

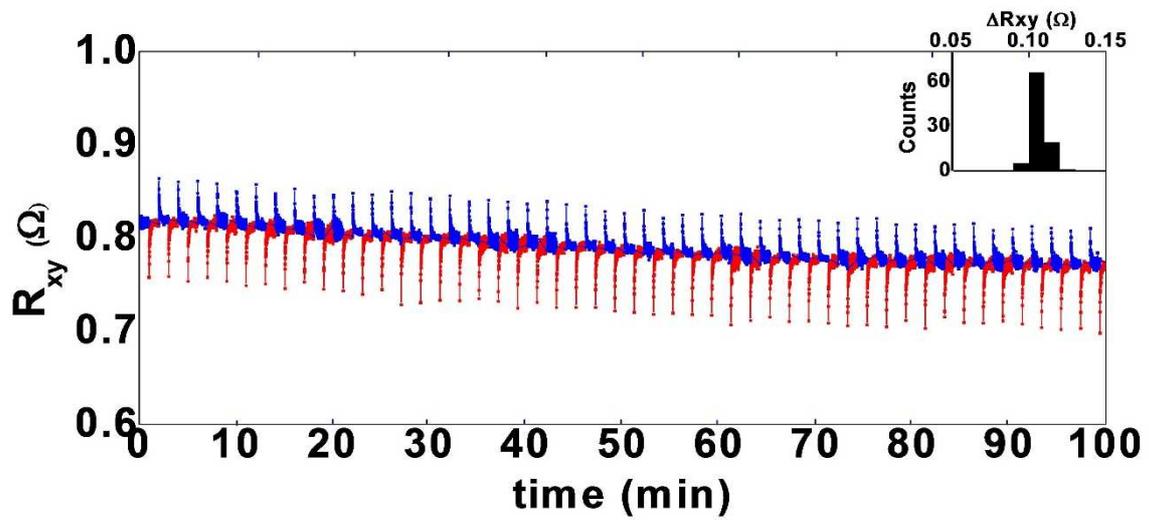

FIG. 3: Orthogonal switching using train of 5 pulses of current 29mA and length 4 ns for axis [1 $\bar{1}$ 0] (blue) and [$\bar{1}$ $\bar{1}$ 0] (red). Inset: histogram showing distribution of readout signal $\Delta R_{xy}$.

**Figure 4**

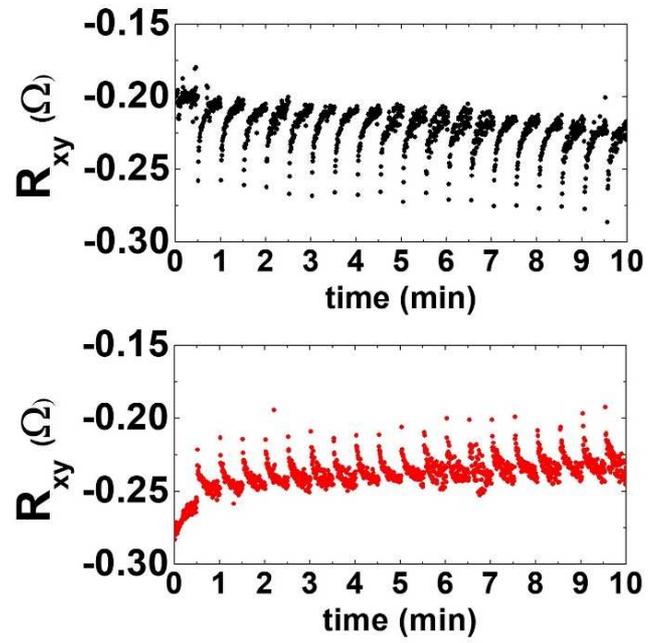

FIG. 4: Plot showing orthogonal switching in a CuMnAs/GaP with a single pulse of current 25mA and pulse width of 4 ns. Top: Current pulses along the [$\bar{1}\bar{1}0$] axis. Bottom: the [$1\bar{1}0$] axis.

Figure 5

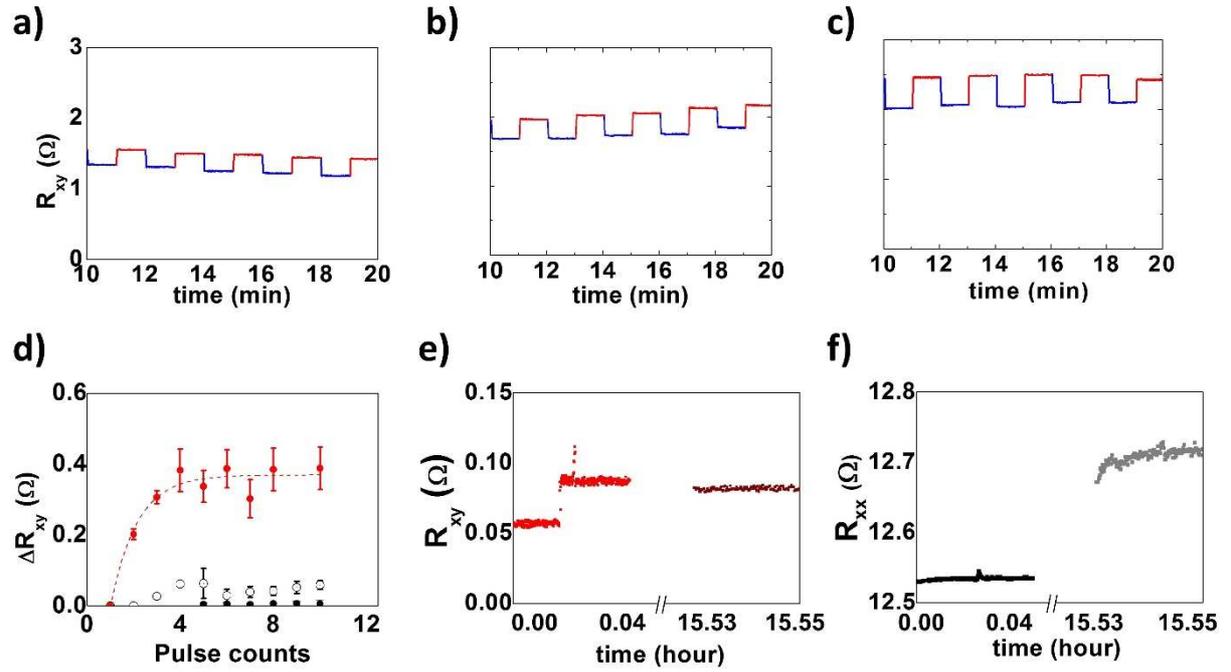

FIG. 5: Polarity switching readout $R_{xy}$ signal across [1 0 0] axis with 2 pulses (a), 3 pulses (b) and 4 pulses (c) of current 29 mA and pulse width of 4ns. (d): Summary plot of number of pulses versus $\Delta R_{xy}$ ($R_{xy}$_high − $R_{xy}$_low) for polarity switching (red) and orthogonal switching (black) and corrected orthogonal switching (black dashed). (e): Plot showing retention of the $R_{xy}$ signal for several hours after polarity pulsing. (f): plot showing the measured $R_{xx}$ over the same time period.

**Figure 6**

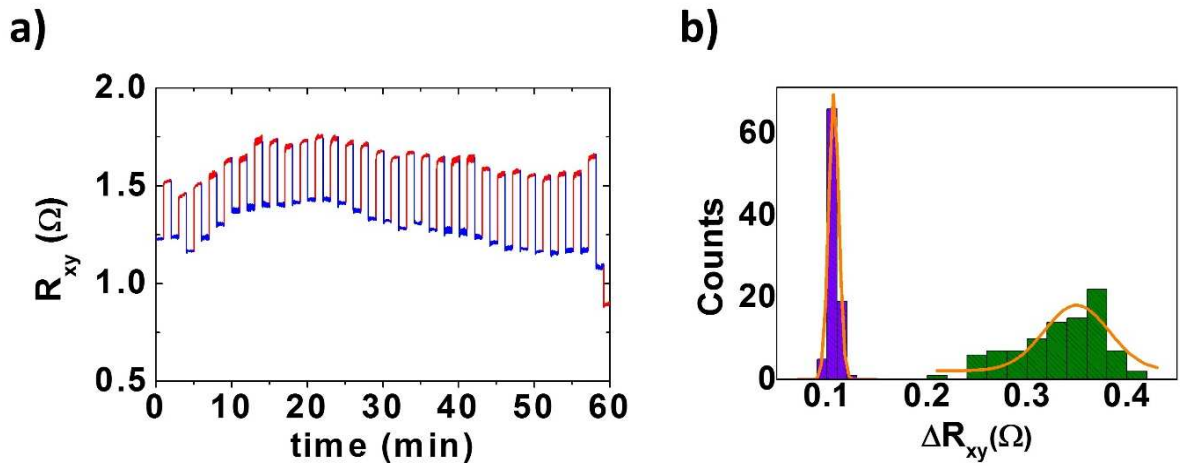

FIG. 6: (a): Output results of $R_{xy}$ signal with 4 ns and 4 pulses indicating polarity switching for 60 attempts across 1 hour. (b): Plot of histograms showing different values of $\Delta R_{xy}$ for orthogonal switching (purple) and polarity switching (green). Orange line is Gaussian fitting with wc (at half maximum) = 0.01 Ω for orthogonal and 0.06 Ω for polarity switching.

**Supplementary Materials**

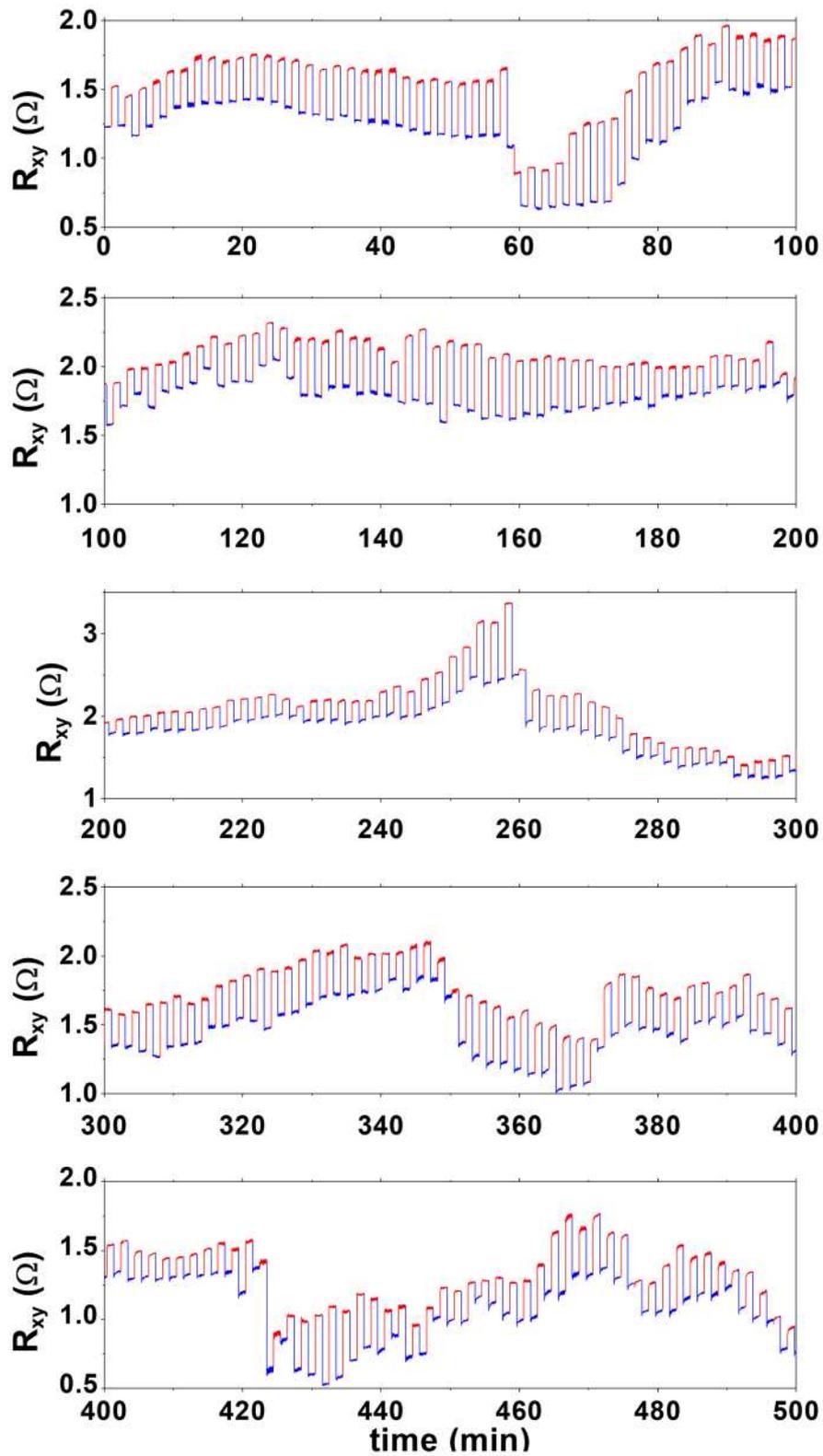

FIG. S1: Output results of $R_{xy}$ signal with 4 ns and 4 pulses across [1 0 0] indicating polarity switching with an interval time of 1 minute between switches for 500 switches. Plots are for intervals of 0 to 100, 100 to 200, 200 to 300, 300 to 400 and 400 to 500 (top to bottom)